# Block Sparse Multi-lead ECG Compression Exploiting between-lead Collaboration

Siavash Eftekharifar[1], Tohid Yousefi Rezaii[2*], Soosan Beheshti[3], Sabalan Daneshvar[2]

1. Centre for Neuroscience Studies, Queen's University, Kingston, ON, Canada
2. Department of Biomedical Engineering, University of Tabriz, Tabriz, Iran
3. Department of Electrical and Computer Engineering, Ryerson University, Toronto, ON, Canada

\* yousefi@tabrizu.ac.ir

**Abstract**: Multi-lead ECG compression (M-lEC) has attracted tremendous attention in long-term monitoring of the patient's heart behavior. This paper proposes a method denoted by block sparse M-lEC (BlS M-lEC) in order to exploit between-lead correlations to compress the signals in a more efficient way. This is due to the fact that multi-lead ECG signals are multiple observations of the same source (heart) from different locations. Consequently, they have high correlation in terms of the support set of their sparse models which leads them to share dominant common structure. In order to obtain the block sparse model, the collaborative version of lasso estimator is applied. In addition, we have shown that raised cosine kernel has advantages over conventional Gaussian and wavelet (Daubechies family) due to its specific properties. It is demonstrated that using raised cosine kernel in constructing the sparsifying basis matrix gives a sparser model which results in higher compression ratio and lower reconstruction error. The simulation results show the average improvement of 37%, 88% and 90-97% for BlS M-lEC compared to the non-collaborative case with raised cosine kernel, Gaussian kernel and collaborative case with Daubechies wavelet kernels, respectively, in terms of reconstruction error while the compression ratio is considered fixed.

## 1. Introduction

Electrocardiography (ECG or EKG) has a substantial role in medical diagnosis since it enables the measurement of the rate and rhythm of the heartbeats. Thus, many anomalies in the heart functionality can be detected by inspecting ECG signals. Long term recording of ECG signals (Sometimes fourteen days or more) is necessary in order to have an accurate diagnosis and detection of certain diseases. Additionally, nowadays, ECG signals are recorded with a very high sampling frequency which results in occupying enormous storage space. Furthermore, with the advent of the new recording tools such as wireless and wearable devices, transmitting ECG signals via Bluetooth/Wi-Fi to another electrical device is of great interest. From the economic point of view, storing, processing and transmitting this vast amount of data are not efficient. Consequently, some compression methods should be applied to ECG recordings in order to reduce the amount of data as much as possible.

A wide variety of compression methods have been used in the past decades. These methods can be divided into three main categories: Direct data compression, transform coding and parameter extraction. Direct data compression, which is based on extracting the important samples of the signal while eliminating the residual, can be considered as the most primary method for ECG compression. Amplitude epoch coding (AZTEC), coordinate reduction time encoding system (CORTES) and scan along polygonal approximation (SAPA) [1-3] are among the direct data compression methods. However, the main shortcoming of these methods is that some of the important samples of the data may be considered redundant and subsequently be deleted during the elimination process. The second approach for compressing ECG signals is transform coding which is based on transforming the signal into another domain using linear transforms such as Fourier and discrete Cosine transforms [4-7]. Nevertheless, the basis functions of these transforms do not resemble the general shape of the ECG signals in the most accurate way and consequently, this issue will negatively affect the compression ratio as well as the reconstruction quality. The last approach which is known as parameter extraction is based on extracting some features of the signal and reconstructing signal using these features. This approach is a combination of the direct and transformation methods. Artificial NN, peak picking and vector quantization are some of the examples of the parameter extraction procedure [8].

Compressed sensing (CS) theory is a novel signal processing technique for efficient acquisition and reconstruction of a signal, by finding solutions to underdetermined linear systems. It is shown that, in comparison with traditional Shannon-Nyquist sampling theorem, CS needs far fewer samples to recover a signal. There are two conditions under which exact recovery is guaranteed. First, the signal must be sparse or approximately sparse in a specific domain. Due to the fact that most of the signals in the nature are not originally sparse, it is necessary to transform them into an appropriate domain in which they have sparse representation. Second, the sensing matrix must satisfy the restricted isometry property (RIP).

In the past few years, CS theory has been applied to compress biological signals including ECG. The single and multi-lead ECG compression using CS theory has gained a lot of attentions in recent years [9-15]. However, in all of these methods the wavelet function (Daubechies kernels) is used to acquire the sparse representation of the signals. A Daubechies wavelet basis can be suitable for most of the signals (including ECGs) in nature because of its shape. Nevertheless, as we have shown in this paper, it is not the best option to model ECGs as sparse signals. In [16] Mcshary et al. have shown that ECG signals can be efficiently modeled by Gaussian functions due to the inherent resemblance of the ECG waves to the Gaussian functions. Inspired by this, in [17], we have used CS theory to compress single-lead ECG signal using a Gaussian dictionary. A comparatively similar idea has also been applied to ECG signals at the same time which is introduced in [18]. Although a Gaussian dictionary seems to be a suitable choice to acquire the sparse representation of the ECG signals, it needs to be an extremely overcomplete dictionary to accurately model the ECG signals, which has an adverse effect on the RIP and coherency condition of the sensing matrix. Furthermore, because the support of the Gaussian functions, is infinite, its tails elongate and sum up, which causes some errors in modeling ECG signals and estimating their sparse coefficients.

In this paper, the CS theory is applied to multi-lead ECG signals. The first step is to transform ECG signals into an appropriate sparsifying domain. To do so, a dictionary with raised cosine kernels is constructed. We show here that raised cosine functions resemble the ECG waves better that Gaussian functions in the sense of having both lower sparsity order and reconstruction error. The choice of raised cosine is due to its finite support compared to Gaussian function, therefore, the tails of this function fall rapidly and diminish more quickly. Furthermore, it can be shown that, using raised cosine kernels instead of Gaussians as the basis function, reduces the redundancy of the dictionary and improves the overall compression performance. We also showed the superiority of the raised cosine wavelet kernel over other traditional wavelet kernels in the simulation results. The raised cosine kernel is also used in [19] where we proposed a method to calculate optimum sparsity order based on minimizing reconstruction error.

It is worth noting that, signals that come from a specific source or several same sources at the same time (for instance same musical instrument that are playing simultaneously) share common properties and have a conspicuous resemblance in their structure. This resemblance in their structures can be exploited in a collaborative manner to improve the compression ratio within CS framework. Different leads of a multi-lead ECG observe and record a same source signal from different locations of the patient's body. Due to this fact, multi-lead ECGs share some common features and properties in their structure. For this reason, while transforming by a sparsifying basis matrix, the sparse models of these signals share a dominant common structure. It is also worth noting that the sparse model for the signal of each lead may also have some innovative components due to the fact that multi-lead ECG signals do not exactly share the same sparsity pattern. This reveals the fact that these signals share common active groups in the dictionary while the sparsity pattern in each group is unique for each signal. This property has been applied in this paper in order to improve the compression performance. After acquiring the sparse representations, a sensing matrix with low coherency which satisfies the RIP condition is created. After obtaining the sparse model for multi-lead ECG, using a suitable sensing matrix the signals can be compressed more effectively. Finally, applying some appropriate recovery method, it is possible to reconstruct the ECG signals with arbitrary small error.

In [20], Sprechmann et al. have proposed a new hierarchical sparse coding framework in a collaborative way, enforcing the same groups for all signals while allowing each group to have its own unique sparsity pattern. The collaborative hierarchical sparse coding framework (C-HiLasso) is used to recover the ECG signal with high precision. Simulation results show that using a raised cosine basis matrix as the dictionary and applying C-HiLasso recovery algorithm improve the compression ratio for a given reconstruction error. Superiority of the proposed method is approved by comparing our method to the results of existing approaches.

The rest of the paper is organized as follows: In Section 2 the mathematical background of CS theory and C-HiLasso reconstruction algorithm are introduced. Section 3 contains the proposed method of constructing a raised cosine basis matrix and the idea of using collaboration and shared group sparsity among ECG signals together with the proper algorithm for the proposed method. The simulation results are presented in Section 4 followed by conclusion remarks which is given in Section 5.

## 2. Background

In this section the mathematical background of CS theory and C-HiLasso are introduced.

### 2.1 Compressed Sensing Theory

The classic Shannon-Nyquist sampling theorem suggests that, in order to avoid deterioration and aliasing during the reconstruction process, the given signal must be sampled by a frequency at least twice greater than the highest frequency existed in the signal. It was a dominant theorem for decades and all applications have been built based on this condition despite its inefficiency in some cases in which the sampling frequency is very high. Around 2004, three mathematicians namely, Emmanuel Candès, Terence Tao, and David Donoho [21-24] challenged this conventional theorem by suggesting that perfect reconstruction may still be possible when the sample-rate criterion is not satisfied, provided that some conditions hold for the underlying signals. They proved that if a signal is exactly or approximately sparse in some appropriate domain, it can be reconstructed with even fewer samples than what the sampling theorem requires. Sparsity means that most of the elements of the signal are zero or nearly zero, leading to a reduced degree of freedom in signal representation. From mathematical point of view a signal $x \in \mathbb{R}^N$ is $k$-sparse when at most $k$ elements of the signal are nonzero. This idea is the basis of CS theory.

Most of the signals in the nature are not sparse in time domain. So, it is necessary to find an appropriate basis matrix and transform the original signal into another domain in which the representation of the signal is sparse. To do so, a suitable basis matrix $\Phi \in \mathbb{R}^{N \times M}$ can be used. In this case, we say a signal $x$ can be referred as a $k$-sparse signal because it can be expressed as $x = \Phi c$ where $c \in \mathbb{R}^M$, $\|c\|_0 \leq k$ and $\|.\|_0$ is the $\ell_0$ norm.

In most cases, signals are simultaneously recorded by multiple sensors/leads placed in different locations. So, the sparse representation of these signals are:

$$X = \Phi C \qquad (1)$$

where $X=(x_1,x_2,...,x_S) \in \mathbb{R}^{N \times S}$ is the matrix of original signals recorded by $S$ sensors, $\Phi=(\varphi_1,\varphi_2,...,\varphi_M) \in \mathbb{R}^{N \times M}$ is the basis matrix and $C=(c_1,c_2,...,c_S) \in \mathbb{R}^{M \times S}$ is the sparse coefficients of each signal where $c_j$ is the sparse representation of $x_j$.

After acquiring the sparse representation of the signals, it is now possible to exploit the CS theory and compress them. The compressed signals can be obtained as;

$$Y=AX \quad (2)$$

where $A \in \mathbb{R}^{m \times N}$ is the sensing matrix and $Y=(y_1,y_2,...,y_S) \in \mathbb{R}^{m \times S}$ is the compressed data matrix corresponding to $S$ original signals. For the sensing matrix we have $m \ll N$, thus the dimensions of the observations (compressed data) are far fewer than the original signals. The value of $m$ must satisfy the following inequality [24]:

$$m \geq u \times k \times \log(N/k) \quad (3)$$

where $u = \frac{1}{2}\log(\sqrt{24}+1) \approx 0.28$.

Combining (1) and (2) the observation signal model can be written as;

$$Y=AX=A\Phi C. \quad (4)$$

Sensing matrix $A$ must satisfy the RIP which in turn guarantees the stable recovery from the compressed signals even if they are contaminated with some random noise.

DIFINITION 1. Matrix $A \in \mathbb{R}^{m \times N}$ satisfies the RIP of order $k$ if there exists a $\delta_k \in (0,1)$ such that

$$(1-\delta_k)\|x\|_2^2 \leq \|Ax\|_2^2 \leq (1+\delta_k)\|x\|_2^2 \quad (5)$$

holds for all $x \in \Sigma_k$.

If a matrix **A** satisfies the RIP of order $2k$, then we can interpret (5) as saying that **A** approximately preserves the distance between any pair of $k$-sparse vectors. This will clearly have fundamental implications concerning robustness to noise [25]. While RIP provides guarantee for recovery of $k$-sparse signals, proving that a matrix satisfies this property is a formidable task and requires a combinatorial search over all $\binom{N}{m}$ submatrices. Thus, in practice it is more convenient to use an equivalent condition, which is easier to compute. The coherence of a matrix is such an equivalent property [26-27]. It is shown that random matrices will satisfy the RIP with high probability if the entries are chosen according to a Gaussian, Bernoulli, or more generally any sub-Gaussian distributions. Consequently, these matrices have small values for their coherence.

*2.2 Collaborative HiLasso reconstruction algorithm*

In the CS framework some appropriate algorithms are needed in order to reconstruct signals from their compressed versions. There exists a wide spectrum of approaches which can be used in the reconstruction process. There are various greedy/iterative algorithms such as orthogonal matching pursuit (OMP) and simultaneous orthogonal matching pursuit (SOMP) which are suitable to use in the CS framework [28-29]. These algorithms are fast and easy to implement. However, when the compression ratio is high, they fail to reconstruct the original signals with desired precision. In other words, when the signals are highly compressed the performance of these algorithms is poor and consequently the reconstruction error increases.

Another powerful approach to recover these signals is $\ell_1$ minimization family. These algorithms solve a convex optimization problem for which there exists an efficient and accurate solution. Collaborative hierarchical lasso algorithm [20] is perfect choice for simultaneously recovering the signals with similar sparsity pattern. This optimization technique is based on the sparse reconstruction by separable approximation (SpaRSA) [30]. The C-HliLasso is an iterative algorithm which solves a sub-problem with a closed form solution at each iteration.

In the SpaRSA framework an algorithm is proposed for solving the unconstraint problems of the general form,

$$\min_{x} \; p(x) := f(x) + \lambda d(x) \quad (6)$$

Where $f \in \mathbb{R}^n \to \mathbb{R}$ is a smooth function, and $d \in \mathbb{R}^n \to \mathbb{R}$, called the regularization function, is finite for all $x \in \mathbb{R}^n$. In the simplest case in CS framework, when the goal is to recover a single signal, (6) can be written as

$$\min_{x} \frac{1}{2}\|y - Ax\|_2^2 + \lambda \|x\|_1 \quad (7)$$

Now if several signals are applied simultaneously, extending the ideas of group lasso and hierarchical lasso which are introduced in [31] and [20], respectively, the C-HiLasso formulation can be obtained. Considering $X$ and $Y$ as the original signals and observations, respectively, the related cost function is given by

$$\min_{X} \; \frac{1}{2}\|Y-AX\|_F^2 + \lambda_2 \Gamma_\varsigma(X) + \lambda_1 \sum_{j=1}^{S}\|x_j\|_1 \quad (8)$$

where $\|.\|_F$ denotes the Frobenius norm, $\Gamma_\varsigma(X) = \sum_{G \in \varsigma}\|X^G\|_F$ and $X^G$ is the sub-matrix formed by all the rows belonging to group $G$ [20]. It can be seen that (8) forces all the signals to share the same groups while the sparsity pattern within the groups depends on the individual signals and differs from the others. It is also worth noting that the values of $\lambda_1$ and $\lambda_2$ will affect the final result. For instance, if $\lambda_2/\lambda_1$ increases, the result is dense inside each group while the group sparsity becomes stronger. However, the optimal values for these parameters are application-dependent and can be found through approaches like cross validation.

**3. The proposed method**

In this section the proposed BIS M-IEC method is introduced. The notion behind modeling multi-lead ECG signals as groups and using the collaboration among leads is also presented. The sparse models are used to more efficiently compress the signals which lead to rather higher compression ratio. The compression process using the proposed method is summarized by a flow chart given in Fig. 1.

### 3.1 Sparse Representation via Raised Cosine Dictionary

As it is previously mentioned, ECG signals are not sparse in time domain, thus it is necessary to use a suitable basis matrix and transform them into an appropriate domain in order to achieve their sparsest representation. Choosing a basis matrix is data and application dependent and it is strongly related to the shape of the signals. In this paper, the raised cosine kernel is used to model ECG signals due to its specific structure and its resemblance to ECG waves. Furthermore, these functions have finite support thus, their tails are shorter and vanish more quickly in comparison with previously used Gaussians kernel. In a Gaussian dictionary, when some Gaussian functions are chosen to model a specific segment of the ECG, their long tails sum up along the time axis, which affect the unintended segments of the ECG period. Thus, the algorithm will choose excessive Gaussian functions in those segments to cancel out these offset errors. This problem deteriorates the modeling process which eventually affects the whole compression performance.

Fig. 2 illustrates the difference between raised cosine and Gaussian kernels. Both kernels have the same shift and scaling parameters. It is evident from Fig. 2 that the tails of the raised cosine function which has finite support diminish more rapidly than the Gaussian function with infinite support.

Furthermore, a raised cosine basis matrix gives a sparser solution for modeling ECG signals than a Gaussian dictionary. The sparsity is defined as;

$$Sparsity\ \% = \frac{N-k}{N} \times 100 \qquad (9)$$

where $N$ is the length of the original signal and $k$ is the number of nonzero entries of the sparse representation. The summary of the results in terms of sparsity for both raised cosine and Gaussian dictionaries is given in Table 1 for three different datasets which are taken from PTB diagnostic ECG database [32]. Letters G and RC stand for Gaussian and raised cosine dictionaries, respectively. C-HiLasso algorithm is used to gain the exact sparse representations. According to (3), a sparser solution leads to rather fewer observations needed to guarantee the unique reconstruction. It is seen that the raised cosine dictionary gives a sparser solution for all 12 ECG leads from different datasets. All in all, it is convenient to construct a raised cosine dictionary in order to acquire the sparse representation of multi-lead ECGs.

To construct a suitable raised cosine basis matrix, the raised cosine functions must be generated in such a way to cover all of the components of the ECG signal within a period. Therefore, it is necessary to generate adequate number of raised cosine functions with different shift and scale parameters. Consider a raised cosine kernel as follows;

$$RC\left(\frac{t-\alpha}{\beta}\right) = \left[1+\cos\left(\frac{t-\alpha}{\beta}\pi\right)\right] \qquad (10)$$

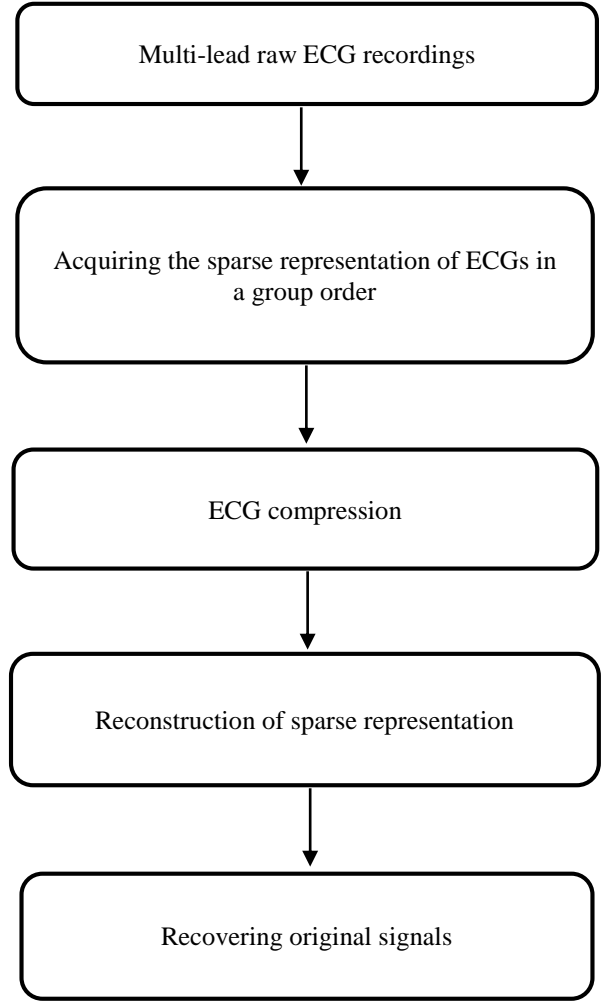

*Fig. 1.* Flow chart of the Multi-lead ECG compression

for $\alpha - \beta \leq t \leq \alpha + \beta$ and zero otherwise. The parameters $\alpha$ and $\beta$ are considered as shift and scale, respectively.

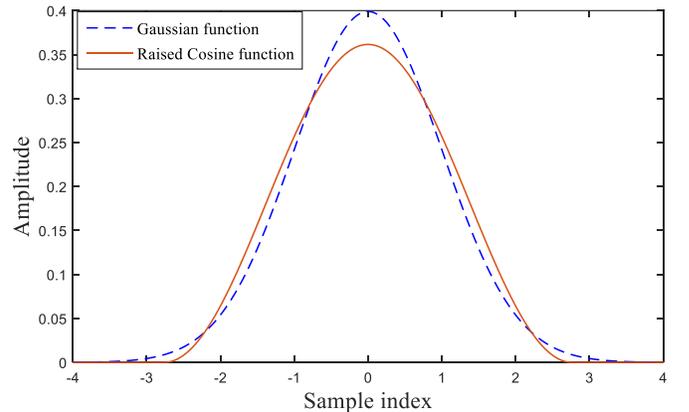

*Fig. 2.* Raised cosine and Gaussian functions of the same shift and scaling parameters.

**Table 1** Sparsity % for All 12 Leads Belonging to Three Different datasets

| Subject No. | | i | ii | iii | avr | avl | avf | v1 | v2 | v3 | v4 | v5 | v6 |
|---|---|---|---|---|---|---|---|---|---|---|---|---|---|
| S0177lrem | RC | 94 | 94.6 | 93.5 | 95 | 94 | 94 | 94.3 | 92.4 | 93 | 94 | 93.8 | 94.3 |
| | G | 90.6 | 88.3 | 91.13 | 91.2 | 93.36 | 92.4 | 84.2 | 90.7 | 90.9 | 90.7 | 92 | 90 |
| S0190lrem | RC | 94 | 91.7 | 91.6 | 94 | 93.7 | 91.4 | 94 | 93 | 93.2 | 93 | 94.3 | 95.1 |
| | G | 92.5 | 85 | 89.83 | 89.6 | 90 | 89.9 | 92 | 89.5 | 90.2 | 89.7 | 93.3 | 93 |
| S0138lrem | RC | 91 | 93.5 | 92.8 | 93.2 | 92.8 | 93.7 | 93.5 | 93.1 | 94.1 | 93.8 | 95 | 95 |
| | G | 88.2 | 90 | 90.1 | 90.5 | 81.5 | 91.4 | 90.3 | 89.5 | 93 | 89.5 | 90.6 | 89.8 |

In most of the ECG signals the R peak is very conspicuous and easy to detect. For instance, consider a period of a healthy and normal ECG signal with a sampling frequency of 1000 Hz. The maximum durations of P-R and R-T intervals in a normal and healthy heartbeat are 200ms and 450ms, respectively. So, after detecting the R peak and according to the mentioned sampling frequency, the maximum number of samples for these segments are 200 and 450, respectively. Therefore, the shifts of the raised cosine kernels which are supposed to cover a full heartbeat must reside in [$S_R$-200, $S_R$+450] where $S_R$ is the sample index of the R peak. Considering this interval, it can be guaranteed that raised cosine kernels cover the whole heartbeat. Furthermore, according to [16] an appropriate interval for scale parameter of the Gaussian kernels is empirically shown to be [0.02-0.6]. Due to the similarity between Gaussian and raised cosine functions, the same interval can be used for scale parameters of the raised cosine kernels. Thus, the basis matrix can be constructed as

$$\Phi=[\varphi_1, \varphi_2, ..., \varphi_M] \quad (11)$$

where

$$\varphi_i = [C(\mathbf{t}, \alpha_i, \beta_i)] = \begin{bmatrix} C(t(0), \alpha_i, \beta_i), C(t(1), \alpha_i, \beta_i), ... \\ , C(t(N-1), \alpha_i, \beta_i) \end{bmatrix}^T,$$

$i=1...,M$, $\mathbf{t}=[t(0),...,t(N-1)]$.

### 3.2 C-HiLasso Algorithm and Group Sparsity for Multi-lead ECG signals

After constructing the raised cosine basis matrix, it is possible to obtain the sparse representation of the ECG signals. There exist a variety of approaches which can be used in this part. However, choosing a proper algorithm is highly important and strongly related to the applied strategy. In numerous applications, we expect certain correlation among a collection of signals (like multi-lead ECGs). These signals share the same components of the dictionary and their support sets are usually similar. In this paper, the goal is to exploit the advantage of the common structure among different ECG leads. The motivation behind this idea is that the multiplication of the sensing matrix and the over-complete dictionary results in a matrix with relatively high coherency due to its similar columns.

In this case, the solutions of the reconstruction algorithm are unstable. In order to reduce the flexibility of the sparse model and increase its stability, it is necessary to choose an algorithm that can appropriately apply the idea of the group sparsity and organize the sparse coefficients in a group order. To do so, the SpaRSA algorithm is used as the core of the recovery procedure and the concept of collaboration among ECG leads is exploited which leads us to avail C-HiLasso reconstruction algorithm. Exploiting this group structure among multi-lead ECGs, improves the performance of the compression process.

In order to elaborate on the between-lead correlation for group sparse models, consider for instance the QRS complex during the depolarization of the ventricles. In this stage, all ECG leads record some meaningful variations in terms of potentials (however with different shapes) in the corresponding time segment; therefore, the same group is activated for all leads corresponding to specific columns of the basis matrix in the mentioned time segment. However, since these variations are not exactly the same, the sparsity pattern inside each group is unique for each lead. In other words, for modeling this part of the ECG signal, a group of columns of the dictionary which happen to be near each other, are chosen by the algorithm. This group is common among all leads but the active columns might be different for each lead inside that group. Figs. 3(a)-(d) clearly illustrate this idea within the QRS complex and its activated group and coefficients are shown in ovals for four leads. This process will be repeated for each wave in an ECG signal. Finally, as the result, we have a sparse representation with the same active groups for each lead, while the active coefficients inside groups are signal dependent and different for each lead. This is the intuition behind using the C-HiLasso algorithm to model multi-lead ECG signals.

After modeling ECG signals and obtaining their sparse representations, the compressed version of the signals can be calculated via (4). However, A$\Phi$ does not satisfy the RIP property with tight bounds, but using the raised cosine kernels reduces the number of columns of the basis matrix which eventually decreases the coherency of the this matrix. The simulation results show that it is possible to use A$\Phi$ in the CS reconstruction and still have good recovery performance for the ECG signals with a very low values for reconstruction error and percentage-root-mean square difference (PRD) also a very low diagnostic distortion based on weighted diagnostic distortion measure (WDD).

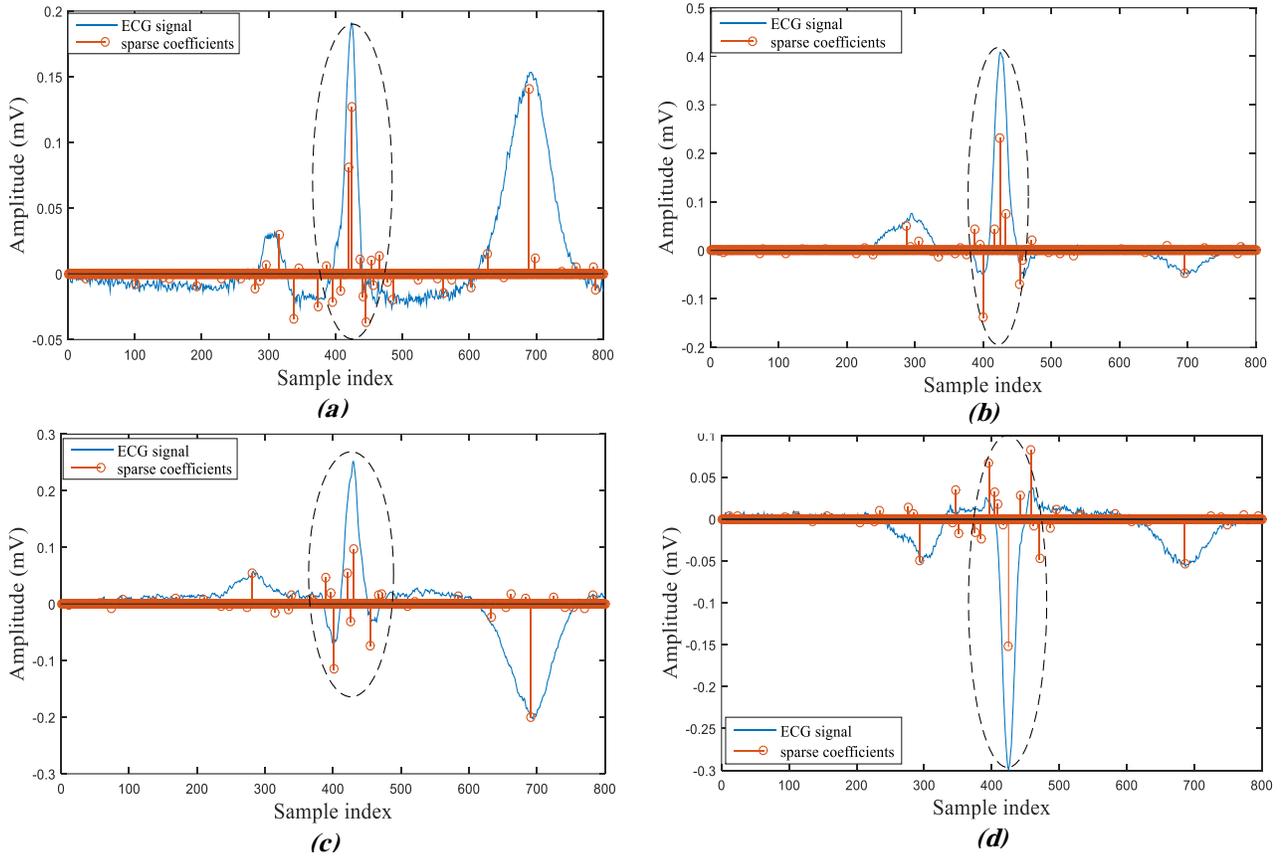

***Fig. 3*** *QRS complex of four ECG leads from s0177lrem dataset and their active groups and coefficients, **(a)** lead i, **(b)** lead ii, **(c)** lead iii and **(d)** lead avr*

The same algorithm (C-HiLasso) is used to recover the ECG signals from the observations.

It is also worth noting that, in addition to preserving the important and main components of the ECG signals after reconstruction, this procedure shows considerable performance in noise reduction of the original raw and noisy signals. This is due to the fact that the sparse coefficients of the ECG signal reflect the main components of the ECG leads and consequently, the noise components within the signal are automatically eliminated. Algorithm 1 shows the summary of BlS M-lEC method.

## 4. Simulation Results

In this section, the simulation results of the proposed method of compressing ECG signals based on sparsifying raised cosine basis matrix, randomly constructed sensing matrix and C-HiLasso reconstruction algorithm are presented. In order to assess the performance of the compression process, the reconstruction error and percentage-root-mean square difference (PRD) are computed for the original and reconstructed signals. We also computed the weighted diagnostic distortion (WDD) to measure the diagnostic quality of reconstructed signals. Finally, the proposed method is compared with the other state-of-the-art approaches. In [9-15] the CS framework is used to compress ECG signals using Daubechies wavelet bases functions. OMP and its variants are among the most popular methods used for reconstruction in the similar literature due to their simplicity; however, they lead to poor performance in term of reconstruction error.

**Algorithm 1**: BlS M-lEC (compression part)

1. **Input:** raised cosine dictionary $\Phi$, twelve lead ECG X
   **Output:** Y, compressed ECG signals
2. **While** stopping criterion is not satisfied **do**
3. Find C, sparse coding of X, using C-HiLasso algorithm and applying raised cosine dictionary $\Phi$.
4. **end while**
5. Creating a suitable sensing matrix A. Entries of A are i.i.d samples form a Gaussian distribution of zero mean and unit variance.
6. Find Y, compressed ECG signal by $Y=AX=A\Phi C$

**Algorithm 1**: BlS M-lEC (reconstruction part)

1. **Input:** Compressed ECG signal Y, sensing matrix A and dictionary $\Phi$
   **Output:** Reconstructed ECG signal $\hat{X}$
2. **while** stopping criterion not satisfied **do**
3. Find reconstructed $\hat{C}$: reconstructing sparse representation from the compressed data Y, using C-HiLasso algorithm and $A\Phi$
4. **end while**
5. Find $\hat{X}$: reconstructed ECG signals using $\hat{C}$ and dictionary $\hat{X}=\Phi\hat{C}$

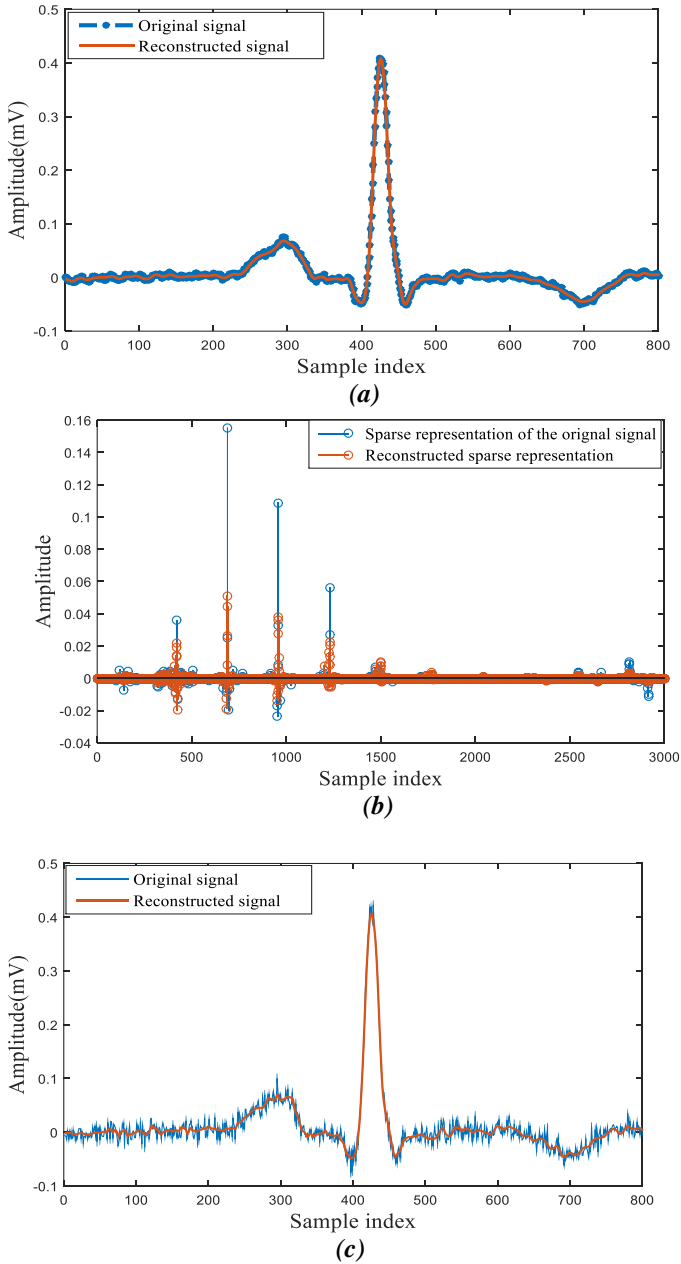

**Fig. 4** *The procedure of ECG signal compression.* **(a)** Original and reconstructed ECG signals **(b)** Original and reconstructed sparse representations **(c)** Original and reconstructed signal with additive Gaussian noise

ECG signals are taken from PTB diagnostic ECG database from the Physionet ATM with sampling frequency of 1000 Hz [32]. The raised cosine basis matrix $\Phi$ is constructed for the signals and C-HiLasso algorithm is used to obtain the sparse representation of 12 leads of the ECG signals. The parameters for C-HiLasso algorithm are chosen empirically to give the best possible results. The entries of the sensing matrix A are assumed to be i.i.d samples from a Gaussian distribution of zero mean and unit variance. The size of matrix A is 130×800. Eventually, the reconstructed sparse signals are obtained using matrix A$\Phi$ by applying the same reconstruction algorithm.

Fig. 4(a) shows one period of the lead ii of the original ECG signal belonging to a 44 years old smoker female. The diagnostic class of this patient is shown in table 2 (S0177lrem). The length of the signal is 800 and the raised cosine basis matrix of the size 800×3000 is used for sparsifying the signal. Afterwards, the sparse representations of the original signals are calculated via (1) which is shown in Fig. 4(b). Then the CS theory is used and the compressed versions of the signals are computed via (2). The sparse representations are reconstructed using the observation matrix by applying the C-HiLasso algorithm. Finally, ECG signals are recovered using the reconstructed sparse coefficients. The reconstructed sparse representation and reconstructed ECG signal of lead ii are depicted in Figs. 4(b) and 4(a), respectively. It is seen that the recovered ECG signal and its sparse representation are nicely fitted to the original ones.

In order to demonstrate the noise reduction property of the proposed method, the white Gaussian noise is added to lead ii of the ECG signal. The SNR is improved by 13 dB and it is seen in Fig. 4(c) that the reconstructed signal is the smoothed and noiseless version of the original one.

In the following, the proposed method is compared with the several other procedures. In order to illustrate the superiority of the proposed method and its compression capability, three performance metrics commonly used in the literature are also used in this section, known as compression ratio, reconstruction error and PRD.

Compression ratio (CR) is used to quantify the reduction in data-representation size produced by a data compression algorithm and is defined as

$$\mathrm{CR} = \frac{\text{Uncompressed size}}{\text{Compressed size}} = \frac{N}{m} \qquad (12)$$

where $N$ and $m$ are the dimensions of the original and compressed signals, respectively. According to (12), it is obvious that CR can vary from 1 to $\infty$ ($1 \le \mathrm{CR} < \infty$).

Reconstruction error is a measure which computes the difference between original and reconstructed ECG signals for all twelve leads defined as

$$\text{reconstruction error} = \frac{1}{S}\sum_{i=1}^{S}\sum_{j=1}^{N}\left(\hat{x}_i(j)-x_i(j)\right)^2 \qquad (13)$$

where $x_i$ and $\hat{x}_i$ are the original and reconstructed signals, respectively and $S$ stands for the number of leads in multi-lead ECG signal.

PRD is another measure for computing the distortion between the matrix of the original signals $\mathrm{X}$ and the matrix of the reconstructed signals $\hat{\mathrm{X}}$ among twelve leads defined as [33]

$$PRD = \frac{\left\|\mathrm{X}-\hat{\mathrm{X}}\right\|_2}{\left\|\mathrm{X}-\overline{\mathrm{X}}\right\|_2} \times 100 \qquad (14)$$

where $\overline{\mathrm{X}}$ is the mean of the original matrix.

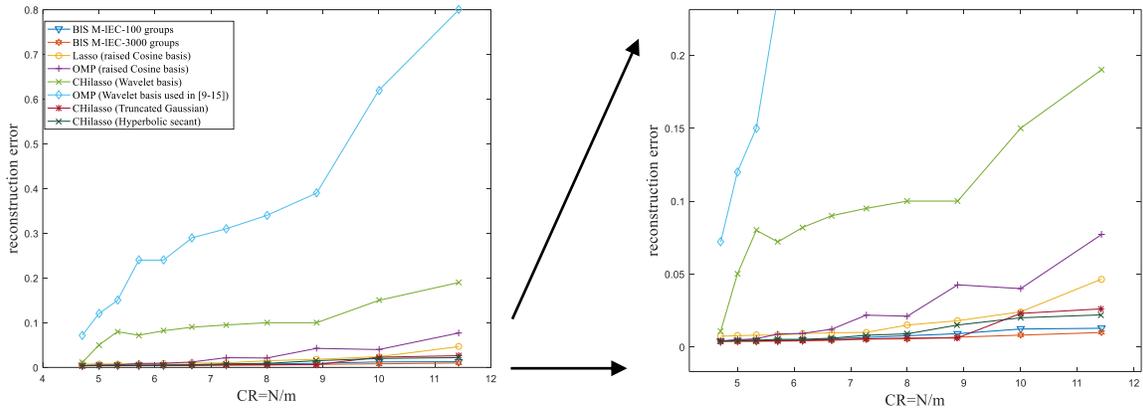

***Fig. 5*** *The reconstruction errors for different methods*

CR versus the reconstruction error as well as PRD for some of the methods are plotted in Figs. 5 and 6, respectively. As it can be seen in these figures, increasing compression ratio results in larger reconstruction error and PRD and the opposite occurs when compression ratio is decreased. It is obvious from Figs. 5 and 6 that the combination of raised cosine basis and C-HiLasso algorithm gives the best result and the lowest reconstruction error and PRD leading to the largest compression ratios in comparison with other methods. As it is mentioned before C-HiLasso treats sparsity in two levels:1) Groups level and 2) sparsity pattern inside each group. It is worth noting that, the number of groups can be manipulated in C-HiLasso algorithm and when it is increased, the reconstruction error and PRD decrease slightly. Considering this point, we have also completed the process when the number of groups is equal to the length of the sparse signal. The results show that the minimum reconstruction error and PRD are achieved when the number of groups is at its maximum but at the expense of more complexity and longer running time.

The reconstruction error and PRD are also computed for other data sets belonging to nine different patients. These signals are taken from PTB diagnostic ECG database from the Physionet database. Table 2 shows the diagnostic classes for ECGs taken from PTP database and a full description of each patients. One of the ECG signals is taken from arrhythmia database (S104) to evaluate the proposed method for arrhythmic ECGs. The point of interest for this signal is located at 3:42s where the signal contains PVC (Premature ventricular complex). PVC's are extra beats which occur from an ectopic focus on the ventricle wall. They may occur for several reasons i.e., diet, fatigue, stress and etc. The summary of the results is given in Tables 3 and 4, while the compression ratio is fixed at 10. Regarding these results, the minimum reconstruction errors for all subjects belong to C-Hilasso with raised cosine basis matrix. The letters RC, W, HS and TG stand for raised cosine, wavelet, Hyperbolic Secant and truncated Gaussian bases functions, respectively. For the wavelet basis the Daubechies 3 (db3), 4 (db4) and raised cosine kernels are used as the sparsifying basis and the expansion is done for 3 levels for each of the kernels.

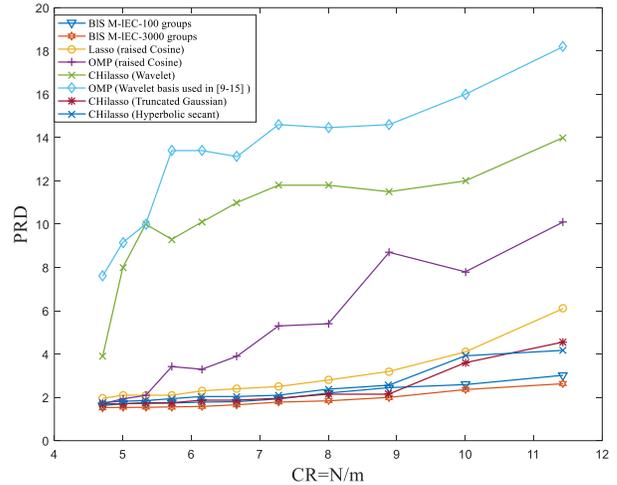

**Fig. 6** *PRD for different methods*

In order to demonstrate the superiority of the raised cosine dictionary over Gaussian, the performance of these dictionaries is also compared in terms of reconstruction error. The reconstruction error for twelve leads of an ECG signal is computed for both dictionaries with different values for *M* and *m*, which are the size of the columns of the dictionary and the length of the compressed signal, respectively. The results suggest that the reconstruction error of the raised cosine dictionary is lower than Gaussian. This difference is more conspicuous in lower values for *m* and *M* which means that using the raised cosine kernels instead of Gaussians, reduces the redundancy of the dictionary and maximizes the CR while having the minimum reconstruction error. The results are shown in Fig. 7.

**Table 2** Diagnostic classes

| Subject no. | Reason for admission | Localization | Additional diagnosis | Smoker |
|---|---|---|---|---|
| S0190lrem | Myocardial infarction | Infero-lateral | Arterial hypertension | Yes |
| S0195lrem | Myocardial infarction | Infero-lateral | No | No |
| S0242lrem | Myocardial infarction | Infero-lateral | M. Bechtere | Yes |
| S0327lrem | Myocardial infarction | Infero-lateral | M. Bechtere | Yes |
| S0031lrem | Myocardial infarction | Anterior | Skoliosis | No |
| S0138lrem | Myocardial infarction | Infero-lateral | Hyperglykemia | No |
| S0550rem | Cardiomyopathy | No | HOC | Unknown |
| S0177lerm | Myocardial infarction | No | No | Yes |

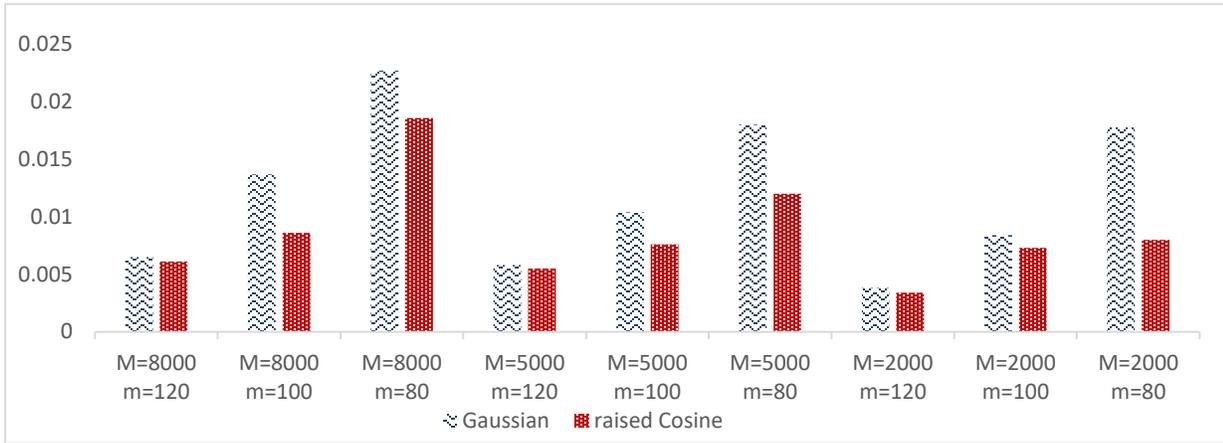

*Fig. 7* The raised Cosine and Gaussian dictionaries reconstruction errors

**Table 3** Reconstruction error using different algorithms and basis functions

| Subject no | BlS M-lEC | Lasso (RC) | OMP (RC) | OMP (W) | C-HL (W-Db 3) | C-HL (HS) | C-HL (TG) | C-HL (W-Db 4) | C-HL (W-RC) |
|---|---|---|---|---|---|---|---|---|---|
| S0190lrem | **0.0054** | 0.0109 | 0.0204 | 1.7 | 0.4554 | 0.0138 | 0.0092 | 0.5076 | 0.0065 |
| S0195lrem | **0.0075** | 0.0157 | 0.0230 | 0.6 | 0.7352 | 0.009 | 0.0102 | 0.8563 | 0.0077 |
| S0242lrem | **0.0277** | 0.0504 | 0.0594 | 0.67 | 0.57 | 0.0423 | 0.0482 | 0.1195 | 0.0285 |
| S0327lrem | **0.0103** | 0.0142 | 0.0416 | 2.1 | 2.7 | 0.0237 | 0.011 | 0.5995 | 0.014 |
| S0031lrem | **0.0136** | 0.0152 | 0.342 | 0.37 | 0.63 | 0.0178 | 0.0166 | 0.1142 | 0.0136 |
| S0138lrem | **0.0064** | 0.01 | 0.0219 | 0.5381 | 0.56 | 0.009 | 0.0201 | 0.1219 | 0.007 |
| S0550rem | **0.0116** | 0.02 | 0.0355 | 2.7 | 4.9 | 0.0267 | 0.0125 | 0.1433 | 0.0122 |
| S0177lrem | **0.0081** | 0.024 | 0.04 | 0.62 | 0.15 | 0.02 | 0.023 | 0.1201 | 0.0087 |
| S104 | **0.005** | 0.007 | 0.013 | 0.04 | 0.035 | 0.008 | 0.0065 | 0.4558 | 0.0055 |
| Average improvement (%) | | **37%** | **75%** | **98%** | **97%** | **39%** | **29%** | **90%** | **8%** |

**Table 4** PRD for Different algorithms and basis functions

| Subject no. | BlS M-lEC | Lasso (RC) | OMP (RC) | OMP (W) | C-HL (W-Db 3) | C-HL (HS) | C-HL (TG) | C-HL (W-Db 4) | C-HL (W-RC) |
|---|---|---|---|---|---|---|---|---|---|
| S0190lrem | **2.7** | 4.67 | 5.03 | 56.8 | 31.56 | 4.85 | 4.7 | 32.38 | 2.7 |
| S0195lrem | **2.9** | 5.6 | 5.6 | 23.47 | 32.61 | 3.49 | 3.87 | 40.56 | 2.99 |
| S0242lrem | **4.5** | 7.3 | 5.8 | 18 | 20 | 6.39 | 6.82 | 11.62 | 4.7 |
| S0327lrem | **3** | 4.9 | 5.3 | 38.88 | 50 | 4.78 | 3.1 | 27.09 | 3 |
| S0031lrem | **5.5** | 5.6 | 6 | 22.82 | 46.7 | 6.43 | 6.5 | 13.09 | 5.8 |
| S0138lrem | **2.8** | 4.3 | 5.4 | 28.62 | 41.2 | 4.72 | 3.9 | 17.49 | 2.85 |
| S0550rem | **2.5** | 2.9 | 3.83 | 35.02 | 65.03 | 3.84 | 2.52 | 9.59 | 2.56 |
| S0177lrem | **2.59** | 4.1 | 7.8 | 16 | 12 | 3.92 | 3.9 | 11.59 | 3.55 |
| S104 | **3** | 5.7 | 13.2 | 81.77 | 75.8 | 8.38 | 8.2 | 92.99 | 3 |
| Average improvement (%) | | **33%** | **41%** | **88%** | **90%** | **35%** | **26%** | **79%** | **4.66%** |

Furthermore, it is known that PRD is not the best measure to compare the original and reconstructed signals in term of diagnostic features. Zigel et al. [34] introduced a new distortion measure for ECG signal compression called weighted diagnostic distortion based on diagnostic features of ECG signals. This measure is based on comparing PQRST complex characteristics of the original and reconstructed signal. We applied WDD to quantitatively measure the reconstruction quality of the proposed method. The diagnostic features of PQRST complex which are chosen to evaluate the proposed methods are the location, durations and amplitudes. Six diagnostic features were used in this paper: $QRS_{dur}$, $P_{dur}$, $QRS_{amp}$, $P_{amp}$, $T_{amp}$ and $QRS_{sign}$. The following equation is used to calculate WDD;

$$WDD(\beta, \hat{\beta}) = \Delta\beta^T \frac{\Lambda}{tr[\Lambda]} \Delta\beta \times 100 \quad (15)$$

where $\beta$ and $\hat{\beta}$ are the vector of the diagnostic features of the original and reconstructed signals, respectively. $\Delta\beta = \beta - \hat{\beta}$ is the normalized difference vector and $\Lambda$ is the diagonal matrix of weights. In this paper, all the weights are considered to be one, which means that all used features are considered to have the same diagnostic value. For duration and amplitude features, the distance is defined as below according to [34];

$$\Delta\beta = \frac{|\beta_i - \hat{\beta}_i|}{\max\{|\beta_i|, |\hat{\beta}|\}}. \quad (16)$$

The results for WDD measure are given in Table 5. Based on the results, the WDD measure for all subjects is less than 10% which indicates "very good" or "good" quality of reconstruction.

As it is previously mentioned, ECG leads record electrical activity of the heart from different locations on body, thus the variations for each lead happen at the same time. This property makes multi-lead ECGs a good case for algorithms like C-HiLasso. In order to visually demonstrate this property, the pattern of the sparse coefficients of 12-lead ECG signal is shown in Fig. 8. This figure clearly demonstrates the concepts of the same active groups and different innovations for ECG leads. The white bars demonstrate the nonzero coefficients while the black bars represent the zero ones.

## 5. Conclusion

In this paper, a new framework is presented for modeling and compression of multi-lead ECG signals based on CS theory. The between-lead correlation is exploited in order to reduce the overall compression ration while still having desired reconstruction error. The raised cosine kernel function is also used to construct the sparsifying basis matrix which is shown to better track the sparse model of the ECG signal. Then, the C-HiLasso algorithm is used to obtain the block-sparse model of the ECG signal which exploits the correlation among multi-lead signals. Finally, the compressed versions of the signals are achieved by projecting the sparse representations onto the measurement space using the sensing matrix. The efficiency of the proposed method over existing approaches is shown in terms of the reconstruction error and PRD versus CR for both original and reconstructed signals.

**Table 5** WDD% measures

| Subject no | WDD% | Reconstruction quality |
|---|---|---|
| S0190lrem | 1.23 | Very good |
| S0195lrem | 5.81 | Very Good-Good |
| S0242lrem | 4.84 | Very Good-Good |
| S0327lrem | 4.09 | Very Good-Good |
| S0031lrem | 8.31 | Very Good-Good |
| S0138lrem | 10 | Very Good-Good |
| S0550rem | 3.5 | Very Good-Good |
| S0177lerm | 6.54 | Very Good-Good |
| S104 | 2.2 | Very Good |

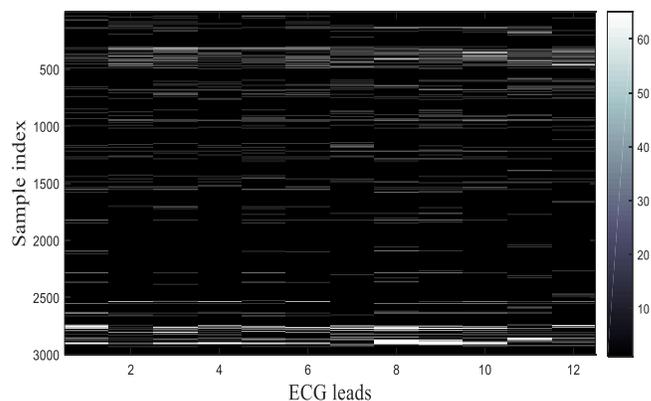

*Fig. 8 Sparse coefficients of 12 leads BlS M-lEC, white bars are the active while the black bars are zero coefficients.*